\documentclass[12pt,aps]{revtex4}

\usepackage[a4paper]{geometry}
\usepackage{amsmath}
\usepackage{amssymb}
\usepackage[dvips]{graphicx}
\usepackage{psfig}

\begin{document}

\begin{center}
{\large \bf
Statics and dynamics of the ten-state nearest-neighbor Potts glass on the
simple-cubic lattice}\\
Claudio Brangian$^1$, Walter Kob$^2$, and Kurt Binder$^1$\\
$^1$ Institut f\"ur Physik,
Johannes-Gutenberg-Universit\"at Mainz, Staudinger Weg 7, \\
D-55099 Mainz, Germany \\
$^2$ Laboratoire des Verres, Universit\'e Montpellier II, F-34095
Montpellier, France

\end{center}

\vspace*{-2mm}
\par
\noindent
\begin{center}
\begin{minipage}[h]{132mm}
We present the results of Monte Carlo simulations of two different
Potts glass models with short range random interactions. In the first
model a $\pm J$-distribution of the bonds is chosen, in the second
model a Gaussian distribution. In both cases the first two moments
of the distribution are chosen to be $J_0=-1$, $\Delta J= + 1$, so
that no ferromagnetic ordering of the Potts spins can occur. We find
that for all temperatures investigated the spin glass susceptibility
remains finite, the spin glass order parameter remains zero, and that the
specific heat has only a smooth Schottky-like peak. These results can be
understood quantitatively by considering small but independent clusters of
spins. Hence we have evidence that there is no static phase transition at
any nonzero temperature. Consistent with these findings, only very minor
size effects are observed, which implies that all correlation lengths
of the models remain very short. We also compute for both models the
time auto-correlation function $C(t)$ of the Potts spins. While in the
Gaussian model $C(t)$ shows a smooth uniform decay, the correlator for the
$\pm J$ model has several distinct steps. These steps correspond to the
breaking of bonds in small clusters of ferromagnetically coupled spins
(dimers, trimers, etc.). The relaxation times follow simple Arrhenius
laws, with activation energies that are readily interpreted within
the cluster picture, giving evidence that the system does not have a
dynamic transition at a finite temperature. Hence we find that for the
present models all the transitions known for the mean-field version of
the model are completely wiped out. Finally we also determine the time
auto-correlation functions of individual spins, and show that the system
is dynamically very heterogeneous.

\end{minipage}
\end{center}

\vspace*{0mm}
\par


\newpage

\section{Introduction}
\label{sec1}
Recently much attention has been given to the study of several models of
mean field spin glasses \cite{1,2,3,4}, such as the spin glasses with
$p-$spin interaction \cite{5,6} and $p-$state Potts spin glasses with
$p > 4$ \cite{1,7,8,9,10,11,12,13,14,15,16,17,18,19}. The interest in
these latter models is related to the fact that they exhibit two distinct
transitions: A dynamic transition from ergodic to non-ergodic behavior at
a temperature $T_D$ where the phase space of the systems splits into many
valleys separated by infinitely high barriers of the free energy , and a
static transition at a temperature $T_0 < T_D$. Although at the static
transition a (static) glass order parameter appears discontinuously
and the internal energy and entropy have kink singularities, there
is no latent heat (Fig.~\ref{fig1}). Due to these properties it has
been proposed that these spin glass models are generic models for the
structural glass transition from a supercooled fluid to an amorphous solid
\cite{20,21,22,23,24,25}. In fact, the slowing down of the relaxation
near $T_D$ is described by the same type of equations \cite{9,10,11,15}
as has been proposed before in the framework of the so-called idealized
mode coupling theory (MCT) \cite{26,27,28,29} to describe the onset of
the slowing down in supercooled fluids near the glass transition, and
it is known that MCT describes a variety of experiments and simulations
quite successfully \cite{27,29,30}. In addition, the $T-$dependence of the
entropy of these models (Fig.~\ref{fig1}) is reminiscent of the behavior
of the entropy of supercooled fluids near the glass transition, which
can be extrapolated to the Kauzmann temperature $T_K$ (which is below the
experimental glass transition temperature $T_g$) \cite{31} at which it
seems to be equal to the entropy of the corresponding crystal. Since one
does not expect that the entropy of the (disordered) fluid is less than
the entropy of the (ordered) crystal, one often invokes a static phase
transition from the (metastable) supercooled fluid to a (metastable)
ideal glass at a temperature $T_0 \geq T_K$ \cite{32,33,34}. However,
the existing theories for such a static transition underlying the
structural glass transition rely on mean-field type approximations
whose validity is questionable \cite{35,36}. Since real systems have
short range interactions, it is relevant to ask which features of the
mean field Potts glass (the scenario of Fig.~\ref{fig1} holds true for
infinite range interactions) can also be found in systems with short range
interactions. In fact, one must expect that in a short range system the
temperature $T_D$ can only have the character of a crossover temperature,
where valleys in phase space appear that are separated by barriers of
large but finite height. Such a scenario is also suggested by the extended
form of MCT \cite{28} and is certainly also required from an analysis
of experimental data \cite{29} which clearly rules out a true power-law
divergence of the relaxation time at $T_D$ in supercooled fluids. Instead
one finds in experiments a smooth crossover from an apparent power-law
for $T>T_D$ to an activated behavior (Arrhenius-behavior: $\ln \tau
\propto 1/T$) for $T<T_D$ or an even stronger $T-$dependence. Of course,
if the singularities at $T_D$ are rounded off in systems with short range
interactions it is of interest to ask whether the static transition at
$T_0$ is still observed or is also wiped out by the fluctuations that
are neglected in mean field theories. Eastwood and Wolynes have recently
put forward some arguments by which supposedly it can be decided whether
or not a short range model shows close to $T_0$ the behavior expected
from the mean field theories~\cite{46}. However, these arguments rely
on the idea that it is possible to partition a given configuration into
droplets which have a well-defined free energy and surface tension. To what
extend such ideas are indeed applicable also to simple spin glasses with
nearest neighbor interactions is presently an open question, however.

The motivation for the present study is to address some of the issues
discussed in the previous paragraph by means of numerical simulations. The
systems considered are $d=3$ dimensional $10-$state Potts glasses with
nearest neighbor interactions~\cite{37}. In Sec. \ref{sec2} we will
describe the models that are studied in the present paper and define
the quantities that are calculated. The main effort of our work is
concentrated on the $\pm J$ model, and the results on the static and
dynamic properties of that model will be presented in Secs. \ref{sec3} and
\ref{sec4}.  Finally, in Sec. \ref{sec5} we will present some results on
the Gaussian model, while in Sec. \ref{sec6} we summarize our conclusions.

\section{Models and details on the simulation}
\label{sec2}
In this section we define the systems that are studied in the present
work and introduce the observables that we will consider.  Also some
details on our simulation procedures are given.

The Hamiltonian of the $p$-state Potts glass is given by

\begin{equation}
\mathcal{H}=-\sum\limits_{\langle i,j \rangle} \, \, J_{ij} (p
\delta_{\sigma_i \sigma_j} -1) \quad ,
\label{eq1}
\end{equation}

\noindent
where the ``exchange constants'' $J_{ij}$ are quenched random variables,
and the symbol $\langle i,j \rangle$ means that each pair of nearest
neighbor sites $i,j$ on the simple cubic lattice is included in the
sum only once. In the present work we will always use $p=10$, and will
set the Botltzmann constant $k_B=1$. Most of our work concentrates
on a bimodal distribution

\begin{equation}
P(J_{ij})= x \delta (J_{ij} -J) + (1 - x) \delta (J_{ij} + J) \quad ,
\label{eq2}
\end{equation}

\noindent
since in a simulation of this model the transition probabilities used in
the Monte-Carlo algorithm for the assignment of a new value of a Potts
spin (we used a single spin-flip type algorithm) can be pre-computed
and stored in a small table. Hence they need not be recomputed in
the course of the simulation, which in turn leads to a faster algorithm.

If we denote the first two moments of the distribution $P(J_{ij})$
as $J_0$ and $(\Delta J)^2$, respectively,

\begin{equation}
J_0=[J_{ij}]_{\textrm{av}}, \qquad (\Delta J)^2=[J^2_{ij}]_{\textrm{av}} -
([J_{ij}]_{\textrm{av}})^2 \quad,
\label{eq3}
\end{equation}

\noindent
we obtain the relation with $x$ and $J$ from Eq.~(\ref{eq2}) as

\begin{equation}
x=\frac{1}{2}(1+ J_0/J) \quad  , \quad J = \sqrt{J_0^2 + (\Delta J)^2} \quad .
\label{eq4}
\end{equation}

\noindent
(Note that the brackets $[.]_{\textrm{av}}$ in Eq.~(\ref{eq3}) stand for
the average over the distribution $P(J_{ij})$ from Eq.~(\ref{eq2}). In
practice this was done by using typically 100 independent realizations
of bond configurations.)

To fix the temperature scale we set $\Delta J= 1$ and furthermore we
choose $J_0=-1$ to avoid the tendency to ferromagnetic ordering. Note that
in the Ising spin glass~\cite{1} the choice $J_0=0$ is most natural and
does not lead to any ferromagnetic correlations in that model. However,
this is no longer true in Potts glasses with $p>3$ as we will show now:
Consider the ferromagnetic susceptibility

\begin{equation} 
\chi_{\textrm{FM}}=N [\langle m^2 \rangle ]_{\textrm{av}} \quad,
\label{eq5}
\end{equation}

\noindent
where $m$ is the ``magnetization'' per spin of the Potts model (using
the simplex representation discussed below) and $N=L^3$ is the number
of sites in the $L \times L \times L$ simple cubic lattice. (Note
that in order to minimize finite size effects we always used periodic
boundary conditions.) The brackets $\langle . \rangle$ stand for the
usual canonical average which were calculated by standard Monte Carlo
averaging \cite{38,39}, using runs of a typical length of $10^8$ Monte
Carlo steps (MCS) per site. (Before these production runs were started,
we equilibrated the samples for the equal amount of time.)

The temperature dependence of $\chi_{\textrm{FM}}$ for the case
$J_0=0$ and $J_0=-1$ is shown in Figs.~\ref{fig2}a and \ref{fig2}b,
respectively. For $J_0=0$, $\chi_{\textrm{FM}}$ gets rather large at
low temperatures and increases rapidly with increasing $L$. In contrast
to this the curves for $J_0=-1$ stay on the order of unity in the whole
temperature range and show, within the statistical errors, no systematic
dependence on the lattice size. Hence we conclude that the choice $J_0=-1$
does not lead to any ferromagnetic ordering of the system.

According to Eq.~(\ref{eq4}) the choice $\Delta J=1$ and $J_0=-1$ leads to 

\begin{equation} 
x= (1-1/\sqrt{2})/2 \approx 0.146 \quad \mbox{and} \quad J= \sqrt{2},
\label{eq6}
\end{equation}

\noindent
and these are the values that we use in this work.

In addition to the $\pm J$ model defined by Eqs.~(\ref{eq1}) and
(\ref{eq2}), we also investigate the Gaussian model defined by

\begin{equation}
P(J_{ij}) = \frac{1}{\sqrt{2 \pi}(\Delta J)} \, \, 
\exp \left[-\frac{(J_{ij} - J_0)^2} {2(\Delta J)^2} \right] .
\label{eq7}
\end{equation}

\noindent
Results for this model will be discussed in Sec. \ref{sec5}.

The static quantities determined in our simulations include the energy
per spin and the specific heat calculated from energy fluctuations,

\begin{equation} 
e(T) =\frac{1}{N}[ \langle \mathcal{H} \rangle]_{\textrm{av}} \quad \mbox{and} 
\quad  c(T) = \frac{1}{NT^2} \left\{[\langle \mathcal{H}^2 \rangle]_{\textrm{av}} 
- [\langle \mathcal{H} \rangle^2 ]_{\textrm{av}} \right\}  \, .
\label{eq8}
\end{equation}

For defining observables like the magnetization, the glass order
parameter, and the time-dependent spin auto-correlation function,
etc., it is useful to choose a representation for the spins that
takes into account the symmetry between their $p$ possible states.
This can be achieved by the so-called ``simplex representation''
\cite{40,41}, in which each of the $p$ states corresponds to a
$(p-1)$-dimensional vector $\vec{S}_\lambda$ pointing towards
the corner of a $p-$simplex, i.e.

\begin{equation} 
\vec{S}_\lambda \cdot \vec{S}_{\lambda'} =(p \delta_{\lambda
\lambda'}-1) \quad \textrm{with} \, \lambda, \lambda' \in \{1,2, \cdots, p\} .
\label{eq9}
\end{equation}

\noindent
To define a spin glass order parameter, we follow the standard method
used for Potts glasses \cite{16,17,18,19,42} by considering two real
replicas $\alpha$ and $\beta$ of the model system, i.e.  two systems
that have identical bond configurations, and to carry out for each of
them an independent Monte Carlo simulation. The order parameter tensor
is then defined as

\begin{equation}
q^{\mu \nu}=\frac{1}{N} \, \sum\limits_{i=1} ^{N} (\vec{S}_{i,
\alpha})^\mu (\vec{S}_{i, \beta})^\nu \quad \mu, \nu \in \{1,2, \cdots,
p-1\} \quad .
\label{eq10} 
\end{equation}

\noindent
In an equilibrium simulation of a finite system with no external
fields that couple to components of the order parameter, there is
no symmetry breaking, and hence it is sufficient to consider the root
mean square order parameter \cite{16,17,18,19,42}:

\begin{equation}
q= \left[\frac{1} {p-1} \, \sum\limits_{\mu, \nu=1}^{p-1} (q^{\mu
\nu})^2 \right] ^{1/2}.
\label{eq11}
\end{equation}

\noindent
By calculating a histogram of $q$ (i.e. one always takes first the thermal
average and then the sample average over the disorder) we can estimate
the distribution $P(q)$ of the order parameter. The various moments of
this distribution are then given by

\begin{equation}
[\langle q^k \rangle ]_{\textrm{av}} = \int\limits_0^1 q^k P(q) dq \quad .
\label{eq12}
\end{equation}

\noindent
The spin glass susceptibility can now be defined in terms of the
second moment:

\begin{equation}
\chi_{\rm SG} = \frac{N} {p-1} [ \langle q^2 \rangle ]_{\textrm{av}}
\quad .
\label{eq13}
\end{equation}

\noindent
Further quantities that exhibit interesting behavior if a phase transition
occurs are the reduced fourth-order cumulant \cite{16,17,18,19,42}

\begin{equation} 
g_4(N,T)= \frac{(p-1)^2}{2} \left( 1 + \frac{2}{(p-1)^2} - 
\frac{[\langle q^4 \rangle]_{\textrm{av}}}{[\langle q^2 \rangle]^2_{\textrm{av}}} \right)
\label{eq14}
\end{equation}

\noindent
and the so-called Guerra parameter \cite{43,marinari98}

\begin{equation} 
G(N,T) =\frac{[\langle q^2 \rangle^2 ]_{\textrm{av}} - [ \langle q^2
\rangle ]^2_{\textrm{av}}} {[ \langle q^4 \rangle ]_{\textrm{av}} - [
\langle q^2 \rangle ]^2 _{\textrm{av}}} \quad .
\label{eq15}
\end{equation}

\noindent
Both quantities are defined such that for $N \rightarrow \infty$
they are zero throughout the disordered phase, while they differ from
zero in the ordered phase. If the system has a phase transition at
a nonzero temperature $T_c$, one expects that the curves $g_4(N,T)$, or
$G(N,T)$, plotted vs. $T$ for different choices of $N$ intersect at
a common intersection point at $T=T_c$. In particular, $G(N,T)$ is a
measure for the lack of self-averaging \cite{43}.

To study the dynamical properties of the system, we calculate the
time auto-correlation function of the Potts spins,

\begin{equation} 
C(t)= \frac{1} {N(p-1)} \, \sum\limits_{i=1}^{N}
[\langle \vec{S}_i (t') \cdot \vec{S}_i (t' + t) \rangle
]_{\textrm{av}} \quad .
\label{eq16}
\end{equation}

\noindent
(Note that in equilibrium, the right hand side of this equation depends
only on the time difference $t$.)

It is also of interest to study a time correlation function that is
analogous to $C(t)$, in which, however, one considers only the time
auto-correlation function of the spin for each lattice site $i$, 
since such a quantity will yield information on
the dynamical heterogeneity \cite{44,45,richert02} of our model system. Below we
will investigate these functions $C_i(t)$ in detail.

To propagate the system in configuration space, we used the heat
bath algorithm \cite{38}.  I. e., to every possible state $\ell$ of a
Potts spin $( \ell \in \{1,2, \cdots, p\})$ a probability $p_{\ell}$
is assigned,

\begin{equation}
p_{\ell} =\frac{\exp (- E_\ell/T)}{\sum\limits_{m=1}^{p} \exp (-E_m /T)}
\label{eq17}
\end{equation}

\noindent
where $E_{\ell}$ is the total energy of the system if the considered Potts
spin is in state $\ell$. At each time step a random spin $i$ is picked,
the probabilities given by Eq.~(\ref{eq17}) are determined, and the
spin is flipped to state $\ell$ with probability $p_{\ell}$. (Hence the
algorithm is quite efficient in finding energetically desirable states even for
large values of $p$, in contrast to the standard Metropolis algorithm.)
Note that in the present case the computation of $E_{\ell}$ involves
only a consideration of the $z=6$ nearest neighbors, unlike the mean
field model studied previously {\cite{16,17,18,19}.

\section{Static properties of the $\pm J$ Potts glass}
\label{sec3}
In Figs.~\ref{fig3} and \ref{fig4} we show the temperature dependence
of the energy per spin $e(T)$ and the specific heat per spin $c(T)$,
Eq.~(\ref{eq8}). We see that for $T \lesssim 2$ the energy seems to become
independent of $T$, i.e. it must be close to the ground state energy
$e(0) \approx -8.8$. This value for $e(0)$ is not far from the value
that one would predict if only the energetically favored ferromagnetic
bonds were fully satisfied. In fact, in such a case we would have (note
the normalization of $\mathcal{H}$ in Eq.~(\ref{eq10}))

\begin{equation}
e(0)= -3 [Jx(p - 1) -J(1-x) (-1)]= -3J[xp-2x+1] \approx -9.2.
\label{eq18}
\end{equation}

\noindent 
This estimate is only $5 \%$ lower than the actually observed value,
showing that the effect of frustration in the present model is very small,
unlike the case of the $\pm J$ Ising spin glass \cite{1}. Since this
frustation is so low, it should be possible to describe the thermodynamic
behavior of the system as a superposition of small independent clusters
of spins. In the Appendix we present such an approach and give the
$T-$dependence of the energy and the specific heat (Eqs.~(\ref{eqa3}),
(\ref{eqa4}), and (\ref{eqa6})). The result of this calculation is
included in Figs.~\ref{fig3} and \ref{fig4} as well. We see that for
the energy the theoretical curve at low $T$ is somewhat lower than the
data from the simulations (since not all effects of the frustration are
taken into account) but that for intermediate and high temperatures the
agreement is very good.

Figure~\ref{fig4} shows that for the case of the the specific heat
the agreement between the data of the simulation and the theoretical
calculation is even better. Note that the shape of $c(T)$ resembles the
one expected for a Schottky anomaly and a closer look at Eq.~(\ref{eqa4})
shows that the full theoretical expression is indeed very similar to such
a functional form (which is $c(T)\propto \beta^2 \,\exp (-pJ/T)$). It is
found that the dominant contribution to $c(T)$ does come from clusters
in which exactly two spins are coupled ferromagnetically and hence for
which the energy needed to flip one spin is $pJ$.

Note that using the theoretical approach presented in the Appendix it
is also possible to calculate the ferromagnetic susceptibility shown in
Fig.~\ref{fig2}b. We have found, however, that the theoretical predictions
are only in qualitative agreement with the results from the simulations
and that significant quantitative differences exist. The reason for this
quantitative discrepancy seems to be that this type of approach is not
reliable in the presence of anti-ferromagnetic couplings, as a similar
calculation for a pure anti-ferromagnetic model (with no disorder)
has shown.

Note that from Fig.~\ref{fig3} and \ref{fig4} we can also conclude that
within the accuracy of the data the energy as well as the specific heat
are independent of the system size, in contrast to our findings for
the mean-field version of the same model~\cite{18,19}.  Hence we can
conclude tentatively that the system does not show a phase transition
at a finite temperature.

The same conclusion emerges from the moments of the spin glass order
parameter distribution (Figs.~\ref{fig5} and ~\ref{fig6}). In particular
we see that the first moment shows a trivial $N^{-1/2}$ dependence
and that $\chi_{\rm SG}$ shows no sign of a divergence in the whole
temperature regime investigated. At the lowest temperatures $\chi_{\rm
SG}$ seems to become constant, which is evidence that the relevant
$T-$range has been probed. The value of the constant for $T\to 0$
is around 3.0. Since for a short range system $\chi_{\rm SG} (T)$ can
be written as a sum of the spin glass pair correlation function over
all distances, this result again implies that these correlations are
small and very short ranged. For high temperatures the $T-$dependence
of $\chi_{\rm SG}$ is compatible with a $1/T-$dependence (see inset of
Fig.~\ref{fig6}), a functional form that is expected for a system {\it
below} its critical dimension~\cite{binder92}.  Also remarkable is the
complete absence of finite size effects in this quantity, in strong
contrast to the mean field version of the model \cite{18,19}. 

Finally we show in Figs.~\ref{fig7} and \ref{fig8} the temperature
dependence of the cumulant $g_4(N,T)$ \{Eq.~(\ref{eq14})\} and the Guerra
parameter $G(N,T)$ \{Eq.~(\ref{eq15})\} for different system sizes. One
recognizes that with increasing $L$ these quantities converge to zero
for all temperatures, which is again consistent with the complete absence
of a finite transition temperature for this model.

\section{Dynamic properties of the $\pm J$ Potts glass}
\label{sec4}
Having discussed the static properties of the $\pm J$ Potts glass,
we now investigate its dynamical properties.

In Fig.~\ref{fig9} we show the time dependence of the (average)
spin-autocorrelation function $C(t)$, see Eq.~(\ref{eq16}), for all
temperatures investigated. Although the data shown are for $L=10$, we
found that within the accuracy of our data the curves are identical to
the ones for larger system sizes. This can be seen in the inset of the
figure where we compare for $T=1.6$ the curve for $L=10$ with the one
for $L=16$. Note that this absence of finite size effects is in strong
contrast to the results found for the mean field version of the present
model~\cite{18,19}.

From the figure we see that the time-dependence of $C(t)$ is rather
unusual in that at low $T$ the function does not show a two step
relaxation process as in Ising spin glasses, but instead shows {\it
two} plateaus: One with a height of approximately 0.6 and another one
with height 0.1. Note that normally the presence of such a plateau is
associated with the existence of some sort of temporary cage in which
the spins are trapped for a certain time. Thus the fact that we now
see two plateaus might be taken as an indication that there are (at
least) two different type of cages. However, as we will show below,
for the present model the reason for the existence of the plateaus is
rather different, despite their similarity with the plateaus seen in the
mean field version of the same model close to the dynamical critical
temperature $T_D$~\cite{9,10,11,18,19}. Last not least we remark that the
existence of further plateaus cannot be ruled out since within the
accuracy of our statistics and within the temperature range accessible in
our simulation it is not possible to make any reliable statements on this.

In order to measure the lifetimes of these plateaus, we define
relaxation times $\tau_i$ via the conditions

\begin{equation}
C(t= \tau_1)=0.4 , \quad C(t= \tau_2) = 0.08, \quad \mbox{and} 
\quad C(t= \tau_3)=0.05 .
\label{eq20}
\end{equation}

\noindent
Thus, the relaxation times $\tau_i$ are obtained from the intersections
of the relaxation functions $C(t)$ at the various temperatures with the
horizontal dashed straight lines shown in Fig.~\ref{fig9}. The time
$\tau_1$ measures the lifetime of the first plateau, whereas $\tau_2$
and $\tau_3$ measure the one of the second plateau. (The reason for
using two different definitions for the latter will be discussed below.)

Many glass forming systems have the property that their time correlation
functions obey the so-called time-temperature superposition principle
(TTSP)~\cite{27}. This means that if one plots a time correlation function
versus $t/\tau$, where $\tau$ is the $\alpha-$relaxation time, one finds
that the correlators for different temperatures collapse onto a master
curve. The existence of the TTSP is one of the main predictions of the
mode-coupling theory of the glass transition and it follows directly from
the solution of the mode-coupling equations~\cite{27}. Since the latter
are identical to the one derived for the time-dependence for $C(t)$ for
the mean field version of the present model, it is of course of interest
to check to what extend the TTSP holds for the present short range model
as well. That this is indeed the case is demonstrated in Fig.~\ref{fig10}
where we plot $C(t)$ vs. $t/\tau_1$ and vs. $t/\tau_2$. We see that with
this type of plot the curves for the different temperatures do indeed fall
on top of each other, as expected from the TTSP. (We mention that a plot
of $C(t)$ vs. $t/\tau_3$ also leads to a superposition of the curves.)

In Fig.~\ref{fig11} we show an Arrhenius plot of the temperature
dependence of the relaxation times $\tau_i$. From this plot we see
that the $T-$dependence of the times $\tau_2$ and $\tau_3$ is the same
and hence we conclude that the exact definition how we defined these
relaxation times is irrelevant.

Also included in the graph are straight lines which represent fits to
the data. Since these fits give a very good description of $\tau_i$
we conclude that the $T-$dependence of $\tau_i$ is just an Arrhenius law:

\begin{equation} 
\tau_i \propto \, \exp (E_A^{(i)} / T) \quad .
\label{eq21}
\end{equation}

\noindent
The activation energies are $E_A^{(1)}\approx 14.6$ and $E_A^{(2)}=
E_A^{(3)} \approx 28.2$. The interpretation of this observation is in
fact quite simple: From Eqs.~(\ref{eq1}) and (\ref{eq6}) we conclude
that the breaking of one ferromagnetic bond costs an energy $p \sqrt{2}
\approx 14.14 \approx E_A^{(1)}$, while overturning a Potts spin which
is coupled by two ferromagnetic bonds to neighboring spins would require
an energy $2p \sqrt{2} \approx 28.3 $ and this is $E_A^{(2)}=E_A^{(3)}$.

In order to understand this behavior in more detail, it is illuminating
to analyze the relaxation dynamics of {\it single} spins. Note that for
such an analysis it is necessary to simulate the system for a time that
is on the order of $10^3$ larger than the longest relaxation time of the
system, since only then it is possible to obtain a statistics that is
sufficiently accurate. Due to this heavy numerical task, we will present
the following analysis only for one realization of the disorder. However,
the results for two different samples gave qualitatively the same results.

In Fig.~\ref{fig12} we show the time dependence of $C_i(t)$ for three
different temperatures ($T=2.4, 2.0$ and 1.6). Since $L=10$, each plot
has $10^3$ different curves. One sees immediately that the relaxation
dynamics of the system is very heterogeneous in that the typical decay
time of the different curves spans many decades and that this range
increases rapidly with decreasing $T$. We find that about 40\% of the
curves relax to (basically) zero very quickly, i.e. within just a few
MCS. This fast relaxation is the reason for the presence of the first
plateau seen in $C(t)$ at a height of approximately 0.6 $(= 1-0.4)$. For
$T=2.4$ (Fig.~\ref{fig12}a) most other curves decay to zero within a time
of $10^3$ MCS. However, if the temperature is decreased to $T=1.6$, see
Fig.~\ref{fig12}c, we see that the typical relaxation time of these curves
has increased to about $10^4$ MCS. Thus we conclude that the second step
in $C(t)$ is due to the relaxation of these spins. In addition we also
recognize from the figure that some spins relax only on the time scale
of $10^6-10^7$ MCS. Thus the existence of the second plateau in $C(t)$
is related to the presence of these slow spins. We also note that the
presence of dynamical heterogeneities has also already been observed
in a short range Ising spin glass~\cite{glotzer98,ricci00}. However, in
that case no simple interpretation of the results has been given so far.

Apart from the curves $C_i(t)$ that, at low $T$, can be grouped together
in a simple way, there are also curves which show a rather complicated
time dependence in that they may show a fast decay at short times,
one or two plateau(s) at intermediate times, before they decay to zero
at long times. Since the total fraction of such curves is only a few
(3-4) percent, and the number of such curves that relax only on the longest
time scales does not seem to increase with decreasing $T$, we feel that
they are not important for the {\it average} relaxation dynamics of the
system. Therefore we will neglect them in the following.

We have found that the shape of the remaining curves can be described
very well by a simple exponential. In addition the $T-$dependence of
relaxation time $\tau_i'$ of each curve is close to an Arrhenius law
and therefore we can make the following Ansatz:

\begin{equation}
\tau_i'= a_i \, \exp (b_i/T), \quad i \in \{1, \cdots, N \}\quad.
\label{eq22}
\end{equation}

\noindent
A histogram with the distribution of the parameters $a_i$ and $b_i$
is shown in Fig.~\ref{fig13}. Note that since we have determined the
$\tau_i'$ at three different temperatures we can use the combinations
($T=1.6$ and $T=2.0$) as well as ($T=2.0$ and $T=2.4$) to determine
the two parameters $a_i$ and $b_i$. However, as can be seen from
Fig.~\ref{fig13}, the two combinations of temperatures give rise to
very similar histograms which is evidence that the Ansatz~(\ref{eq22})
holds and that the way the parameters $a_i$ and $b_i$ were determined
is irrelevant.

While the distribution of $a_i$ seems to be continuous without any
special features (see inset), the one for $b_i$ shows clearly three
peaks: One at $b_i \approx 0$, one at at $pJ$ and a third one at $2pJ$
respectively. Thus we see that the three different types of groups of
spins identified in Fig.~\ref{fig12} can also be found in this analysis.

It turns out that all these features of the relaxation behavior can even
be understood {\it quantitatively}. First we mention our observation
that all of the spins that relax quickly do not have ferromagnetic bonds.
Recalling that each spin has $z=6$ neighbors and that there are $p=10$ states
at disposal, the probability that a spin has only antiferromagnetically
coupled neighbors is given by $(1-x)^6 \approx 0.39$, which accounts for
the numerically estimated fraction of $40 \%$ of fast relaxing spins. More
general, the probability to have exactly $k$ ferromagnetic bonds out of
$n$ bonds is

\begin{equation} \label{eq23}
P(k,n) =\frac{n!} {k! (n-k)!} \, x^k (1-x)^{n-k} \quad .
\end{equation}

\noindent
So, a first estimate would be to say that a spin with exactly $k$
ferromagnetic bonds relaxes with a barrier given by $kpJ$.  Noting that
$P(1,6) \approx 0.40$, $P(2,6) \approx 0.17$, $P(3,6) \approx0.04$ etc.,
we note that this approach would predict a second plateau at a height
of $0.21$ (resulting from the sum of probabilities with $k \geq 2$), and
also predict a third plateau with a rather large height of about $0.046$.
Since we do not observe such a third plateau it is necessary to refine the
approach: While spins with a single ferromagnetic bond need to overcome a
barrier of $pJ$, the spins with two ferromagnetic bonds do not necessarily
need to overcome a barrier of $2pJ$. E.g., if one of the two spins that
are connected ferromagnetically has itself only one ferromagnetic bond,
it will relax with a barrier $pJ$, and thus can subsequently permit to
the first spin with two ferromagnetic bonds also to relax, overcoming
a barrier of $pJ$ only. The probability to have such an arrangement of
bonds is given by $P(2,6)$, that is the probability for the spin to have
exactly 2 ferromagnetic bonds with its neighbors, times the probability
that one of these two spins has no further ferromagnetic bond with its
remaining 5 neighbors, and which is thus $P(0,5)$. This yields $P(2,6)
\cdot P(0,5) \approx 0.077$, so that the total percentage of spins
relaxing with barriers higher than $pJ$ is lowered to $0.13$, very close
to the value found for the second plateau from Fig.~\ref{fig9}. 
Extending this reasoning also to spins relaxing with a barrier $3pJ$,
one estimates that the height of a third plateau would be around $0.006$,
and this is clearly too small to be observable numerically within the
accuracy of our data. Note also that these numerical estimates completely
ignore the small effects due to the frustrated bonds, which still must be
present, as the estimate for the ground state energy shows. Nevertheless,
these arguments show that the relaxation of the model is to a large extend
accounted for by the behavior of small isolated clusters of correlated spins,
and no collective relaxation associated with either dynamic or static
transition is needed.

\section{The Gaussian Potts glass}
\label{sec5}
We now turn our attention to the case where the couplings between
neighboring spins are not given by a bimodal distribution, but by 
a Gaussian one (see Eq.~(\ref{eq7})). As we will see, this change in the
distribution has important implications for the static and dynamic
properties of the system. All results discussed in this section were
obtained for $L=10$.

In Fig.~\ref{fig14} we show the $T-$dependence of the energy per spin. In
agreement with our finding for the $\pm J$ model, this function does
not show any sign for the presence of a singularity. At low temperatures
we see that (with decreasing $T$) the slope of the curve decreases and
that $e(T)$ approaches smoothly its value in the ground state. This
value is close to -5.49, the energy of the system if one assumes that
all interctions are satisfied, i.e.

\begin{equation}
e_{\rm gs}^{\rm appr} = 3\left(\int_{-\infty}^0  P(J_{ij}) J_{ij} dJ_{ij} -
(p-1) \int_0^\infty P(J_{ij}) J_{ij} dJ_{ij}\right) \quad .
\label{eq22b}
\end{equation}

Here $P(J_{ij})$ is the distribution given by Eq.~(\ref{eq7}). The value
of $e_{\rm gs}^{\rm appr}$ is marked in the figure by the horizontal arrow.

That $T-$dependence of the specific heat is shown in
Fig.~\ref{fig15}. Similar to the case of the $\pm J$ model, we find a
Schottky-like peak at around $T\approx 2$. Since in the Gaussian model
there is no gap in the excitation spectrum, by analogy with Ising spin
glasses \cite{1}, we expect a low temperature behavior $c(T) \propto T$
and $e (T) - e(0) \propto T^2$ rather than the behavior proportional to
$\exp (-pJ/T)$ encountered in the $\pm J$ model. However, since for $T
\leq 1.2$ the relaxation times are already rather large, no attempt has
been made to verify these expectations numerically.

Since the ferromagnetic bonds now occur with a broad spectrum of
energies rather than a single energy $pJ$, the argument put forward in
the previous section for the existence of a plateau does no longer apply. In
Fig.~\ref{fig16} we show the time-dependence of $C(t)$ for all temperatures
investigated and from this graph we see that in this $T-$range there
is no sign for the presence of a plateau. Although for long times and
low $T$ the TTSP seems to hold reasonably well, we have not been able
to determine the functional form of the time dependence of $C(t)$. 

Of course it might be that the absence of the plateau is related to the
fact that we are investigating the system at too high temperatures. In
order to check for this possibility we have quenched 10 independent
sample (i.e. different realisations of disorder) to $T=0.5$ and have
annealed the system for $10^7$~MCS before we started a production run of
$10^7$~MCS. Although this time is certainly insufficient to equilibrate
the systems at this temperature, it can be expected that for times
significanly (O(10)) shorter than $10^7$~MCS the correlation function
is close to the one in true equilibrium.

The correlation function as calculated after the quench is included
in Fig.~\ref{fig16} as well.  We see that even at this low temperature
there is no well defined plateau. Instead the time dependence seems to
be compatible with a logarithmic decay, at least in the time interval
$10 \leq t \leq 10^5$. Hence we have evidence that for this model the
mean field prediction for the existence of a dynamical transition does
not seem to hold.

If we define relaxation times similarly as done previously by requiring
that $C(t)$ reaches a particular value, 

\begin{equation} \label{eq24}
C(t= \tau_1)=0.1 , \quad C(t= \tau_2)=0.05, \quad C(t=\tau_3)=0.02
\end{equation}

\noindent
(horizontal dashed lines in Fig.~\ref{fig16}) we find again that
the $\tau_i$ have an Arrhenius temperature dependence (see Fig.~4 in
Ref.~\cite{37}), but the activation energy seems to increase smoothly
the smaller the constant in Eq.~(\ref{eq24}) is chosen. (We find the
activation energies 20.1, 16.4, and 12.8 for $\tau_3$, $\tau_2$ and
$\tau_1$.) Once more, this behavior is again expected from a system
where one has small clusters (dimers, trimers, etc.) of ferromagnetically
coupled Potts spins, and the breaking of this (continuous spectrum) of
ferromagnetic bonds causes the extended tails of the time-correlation
functions $C(t)$ as seen in Fig.~\ref{fig16}.

\section{Concluding remarks}
\label{sec6}

In this paper we have used Monte Carlo simulations to investigate the
static and dynamic properties of the $10$-state Potts glass model with
nearest neighbor random interactions on the simple cubic lattice. We
have found that in the temperature range investigated there is no sign
for the presence of the {\it dynamical} transition from ergodic to nonergodic
behavior nor for the {\it static} discontinuous glass transition that one finds
in the mean field version of this model. Instead the dynamic transition is
replaced by a very gradual onset of slow dynamics, described by relaxation
times that are compatible with simple Arrhenius laws.

For the case of the $\pm J$ model, the spectrum of relaxation times is
essentially discrete -- namely times $\tau_k \propto \exp (kp J/T)$
with $k=1,2$. These times are related to processes in which a
spin is connected to its nearest neighbors via $k$ ferromagnetic bonds
that have to be broken for the spin to flip. In contrast to this, the
Gaussian model has a spectrum of relaxation times that is continuous,
which reflects the fact that also the values of the ferromagnetic bonds
are distributed continuously.

A further important difference between the $\pm J$ model and the
Gaussian model is that the average spin auto-correlation function of
the latter shows, after a fast drop at short time, a smooth decay to
zero. In contrast to this, $C(t)$ for the $\pm J$ model shows several
steps, the existence of which is related to the discrete spectrum of
relaxation times. The auto-correlation function of individual spins
is for most spins a simple exponential, with a decay time that is
given by one of the relaxation times $\tau_k$ mentioned above. Thus,
although there is a pronounced dynamical heterogeneity in the model,
this heterogeneity is localized on a very small length scale (some
spins have only antiferromagnetic bonds to their neighbors and can
relax fully unhindered, others are in ferromagnetically coupled dimers,
trimers, etc., but the concentration of large clusters is vanishingly
small). Due to the strong localization of these excitations it is
unlikely that a description of dynamical heterogeneity in terms of
mesoscopically large regions (``droplets'' where volume and surface
terms in their free energy compete, etc. \cite{25,46}) is applicable,
at least for the present model. Instead we have shown in the present
work that a description in  terms of small clusters of ferromagnetically
coupled spins in a background of spins to which they are ``coupled''
antiferromagnetically is almost quantitatively accurate. (Note that
this antiferromagnetic coupling has almost no effect on the cluster,
due to the large number of states for the spins that still are available
when a bond between them is antiferromagnetic.)  E.g. this description
allows to explain the $T-$dependence of static quantities like energy,
or the specific heat very well, and thus we conclude that in the present
model the spins are correlated only very weakly. The fact that we are
able to give an almost quantitative description of the static as well
as dynamic properties of the system in the $T-$range considered makes
it also very unlikely that a dynamical or static transition occur at
even lower temperatures, although strictly speaking such a possibility
can of course not be ruled out.

Hence we conclude that the idea that short range Potts glasses may provide
a good model for the structural glass transition problem, exhibiting a
rounded version of a dynamic transition (with a crossover from a power-law
divergence of the relaxation time to an Arrhenius behavior with a very
large barrier) and a (very closely avoided) static transition to a glass
phase at a lower but nonzero temperature can {\it not} be maintained, at least
not for the present model. Instead we think that a better model for the
structural glass transition would be a Potts glass with interactions
that have a somewhat longer range than in the models studied here.

Very recently it has been suggested that such a close analogy between
Potts glasses and structural glasses might exist only if the number of
available states is very large, e.g. $p=10^3$ \cite{46}. However, we do
not see any physical basis for such a claim: The choice $\Delta J =1$,
$J_0 =-1$ in the model, necessary to avoid the tendency for ferromagnetic
ordering, leads to a fraction $x \approx 0.146$ of ferromagnetic bonds,
and for larger values of $p$ an even smaller $x$ is needed~\cite{12}.
However, with such a small fraction of $x$ the properties of the model
will be qualitatively the same for $p=10$ and $p=10^3$. Frustration
effects also cannot change this conclusion - the fraction of frustrated
plaquettes with three ferromagnetic bonds and one antiferromagnetic bond
simply is too small (note that plaquettes with one ferromagnetic bond
but three antiferromagnetic bonds are frustrated in an Ising spin glass,
where $p=2$, but not for the large $p$ values considered here). While
the idea \cite{46} that one should discuss the applicability of the
mean field Potts glass transition scenario to real systems in terms
of a Ginzburg criterion \cite{47} is certainly interesting, it seems
that it is not useful for the extremely short range case of a nearest
neighbor Potts glass. This conclusion is of course not that surprising
since our experience with Ising ferromagnets has shown that the Ginzburg
criterion tells little about the nearest neighbor model either, although
it is useful for understanding properties of Ising models with a large
but finite range of the forces \cite{48}. A study of the $10-$state
Potts glass with random interactions with such medium range could be
illuminating, but clearly is computationally very demanding and hence
was not attempted here. In any case, our study shows that it is rather
difficult to draw analogies between spin-glass-type models and models
commonly used to study the problem of the structural glass transition
(hard spheres, Lennard-Jones, etc.) even if the description of such
models within mean-field looks very similar.  \bigskip

{\bf Acknowledgements:} This work was supported by the Deutsche
Forschungsgemeinschaft (DFG) under grant No. SFB262/D1. We thank the
John von Neumann Institut for Computing (NIC) in J\"ulich for a generous
grant of computing time on the CRAY-T3E.

\appendix*
\section{}
In this appendix we briefly describe the analytical calculations we used
to obtain the (approximate) temperature dependence of the energy $e(T)$
and of the specific heat $c(T)$.

To calculate $e(T)$ we consider a Potts-spin that we will denote by
``0''. This spin is connected to $2d$ neighbors ($d$ is the dimension
of the lattice, i.e. in our case $d=3$). Assume that out of the $2d$
bonds that connect the central spin to its $2d$ neighbors, exactly $k$
are ferromagnetic and that hence $2d-k$ are anti-ferromagnetic. In
the following we will calculate the partition function for such a
``$k-$cluster'' of $2d+1$ spins and subsequently its energy. We then make
the approximation that the Potts Hamiltonian given by Eq.~(\ref{eq1}) is
a sum of {\it independent} clusters (using the distribution $P(J_{ij})$
to determine the frequency with which a $k-$cluster occurs).

Using the notation $\Delta := pJ$, we can write the Hamiltonian of a
$k-$cluster as

\begin{equation}
\mathcal{H}_k=-\Delta \sum_{i=1}^{k} \delta_{\sigma_0 \sigma_i} +
\Delta \sum_{i=k+1}^{2d} \delta_{\sigma_0 \sigma_i} +2J(k-d)\quad .
\label{eqa1}
\end{equation}

\noindent
The partition function can now easily be obtained as

\begin{equation}
Z_k= p [p-1+\exp(\beta\Delta)]^k [p-1+\exp(-\beta\Delta)]^{2d-k}\exp[\beta 2J(k-d)]\quad ,
\label{eqa2}
\end{equation}

\noindent
with $\beta=1/k_BT$. The energy can now be calculated by using the
relation $e_k(T) = -Z_k^{-1} \partial Z/\partial \beta$ and one obtains:

\begin{equation}
e_k(T)= \Delta\left\{ \frac{(2d-k) \exp(-\beta \Delta)}{p-1 +  \exp(-\beta \Delta)}
- \frac{k \exp(\beta \Delta)}{p-1+ \exp(\beta \Delta)}\right\}+2J(k-d) \quad .
\label{eqa3}
\end{equation}

The specific heat $c_K(T)$ can now be calculated using $c_K(T)=\partial
e_K(T)/\partial T$ which gives

\begin{equation}
c_k(T)=\Delta^2\beta^2 (p-1) \exp(-\beta \Delta) 
\left\{ \frac{(2d-k)}{[p-1 + \exp(-\beta \Delta)]^2}
- \frac{k}{[(p-1)\exp(-\beta \Delta)+1])^2}\right\}.
\label{eqa4}
\end{equation}

If in the real system we pick an arbitrary spin and consider its $2d$
nearest neighbors there is a probability $W(k)$ that the first spin has
exactly $k$ ferromagnet bonds and $W(k)$ is given by:

\begin{equation}
W(k)= \frac{(2d)!}{k!(2d-k)!} x^k(1-x)^{2d-k} \quad ,
\label{eqa5}
\end{equation}

\noindent
where $x$ is the concentration of ferromagnetic bonds (see
Eqs.~(\ref{eq2}) and (\ref{eq6})). Thus within the approximation
considered here, the value of an observable $Y$ of the system (i.e. energy
per spin or specific heat) is given by

\begin{equation}
Y=\sum_{k=0}^{2d} Y_k \, W(k) \quad .
\label{eqa6}
\end{equation}

\noindent
Using this relation we have calculated the temperature dependence of
the energy per spin as well as the specific heat that are shown in
Figs.~\ref{fig3} and \ref{fig4}, respectively.

\clearpage
\begin{figure}
\centerline{
\psfig{figure=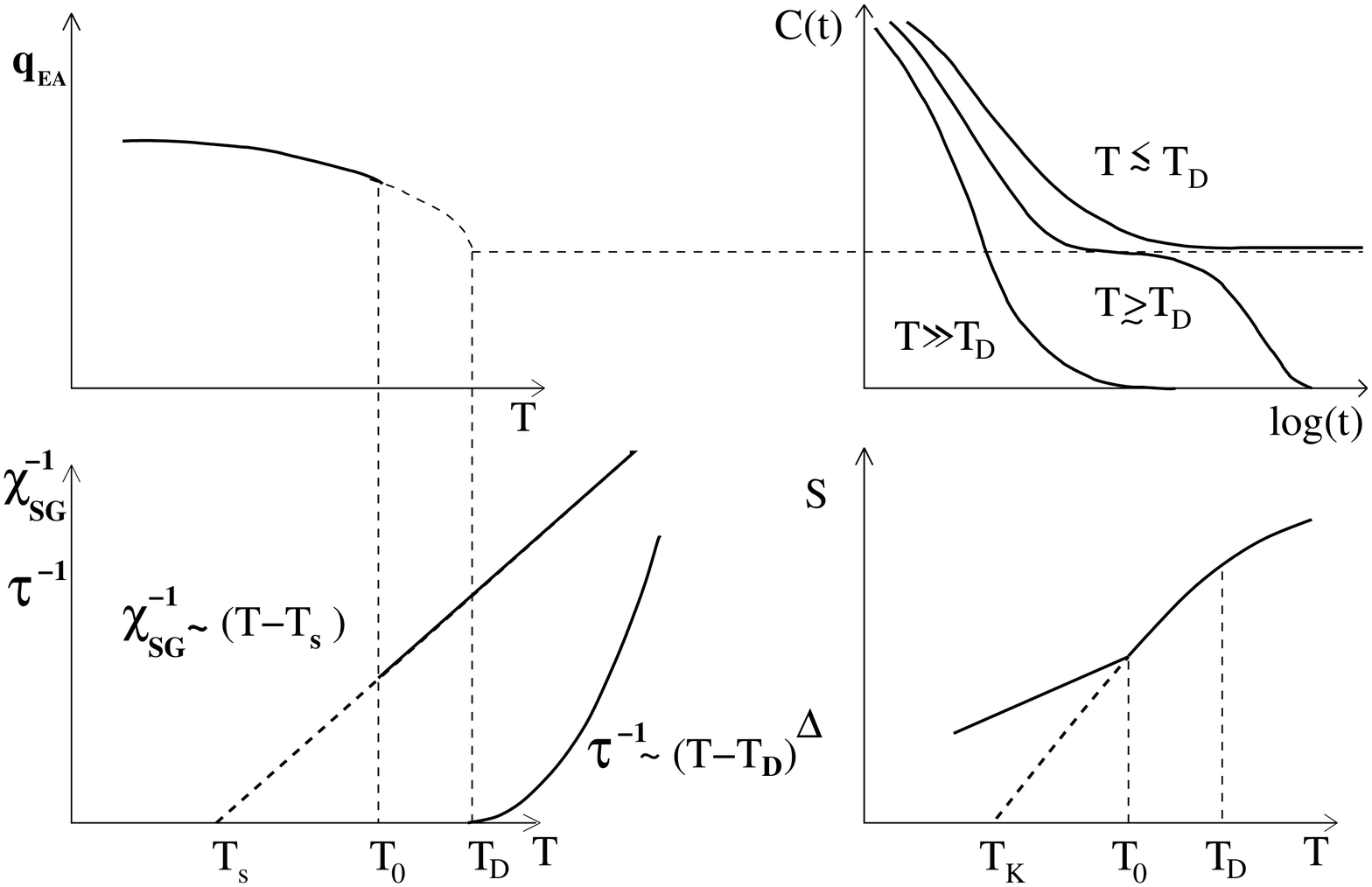,width=11.cm,height=7.0cm}
}
\vspace*{-6mm}

\caption{\label{fig1} Schematic sketch of the thermodynamic and dynamic
transitions in a mean-field $p$-state Potts glass model with $p>4$ states.
The spin glass order parameter $q_{EA}$ is non-zero only for $T<T_0$ and
jumps to zero discontinuously at $T=T_0$. The spin glass susceptibility
$\chi_{\rm SG}$ follows a Curie-Weiss type relation with an apparent
divergence at a ``spinodal temperature'' at $T_s<T_0$, and hence is
still finite at $T_0$. The relaxation time $\tau$ diverges according
to a power-law $\tau \propto (T-T_D)^{-\Delta}$ at a temperature $T_D >
T_0$. This divergence is due to the occurrence of a long-lived plateau
of height $q_{EA}$ in the time-dependent spin auto-correlation function
$C(t)$. The entropy $s(T)$ per spin has no singularity at $T_D$, but
shows a kink singularity at $T_0$ (thus there occurs no latent heat, which
would mean an entropy jump).  The extrapolation of the high temperature
branch of the entropy would vanish at a temperature $T_K \leq T_0$.}
\end{figure}

\newpage
\clearpage

\begin{figure}
\centerline{
\psfig{figure=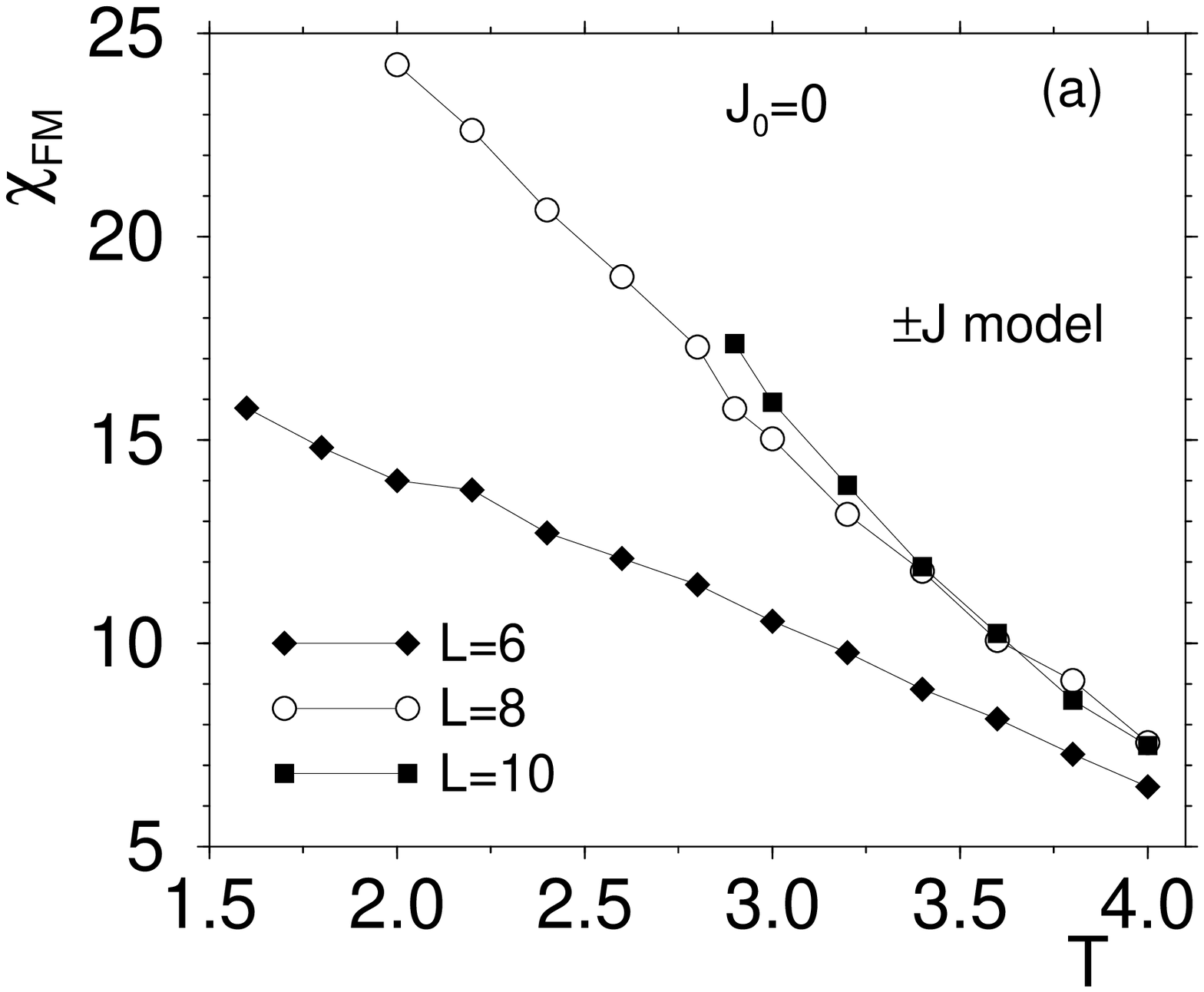,width=8.2cm,height=7.0cm}
\psfig{figure=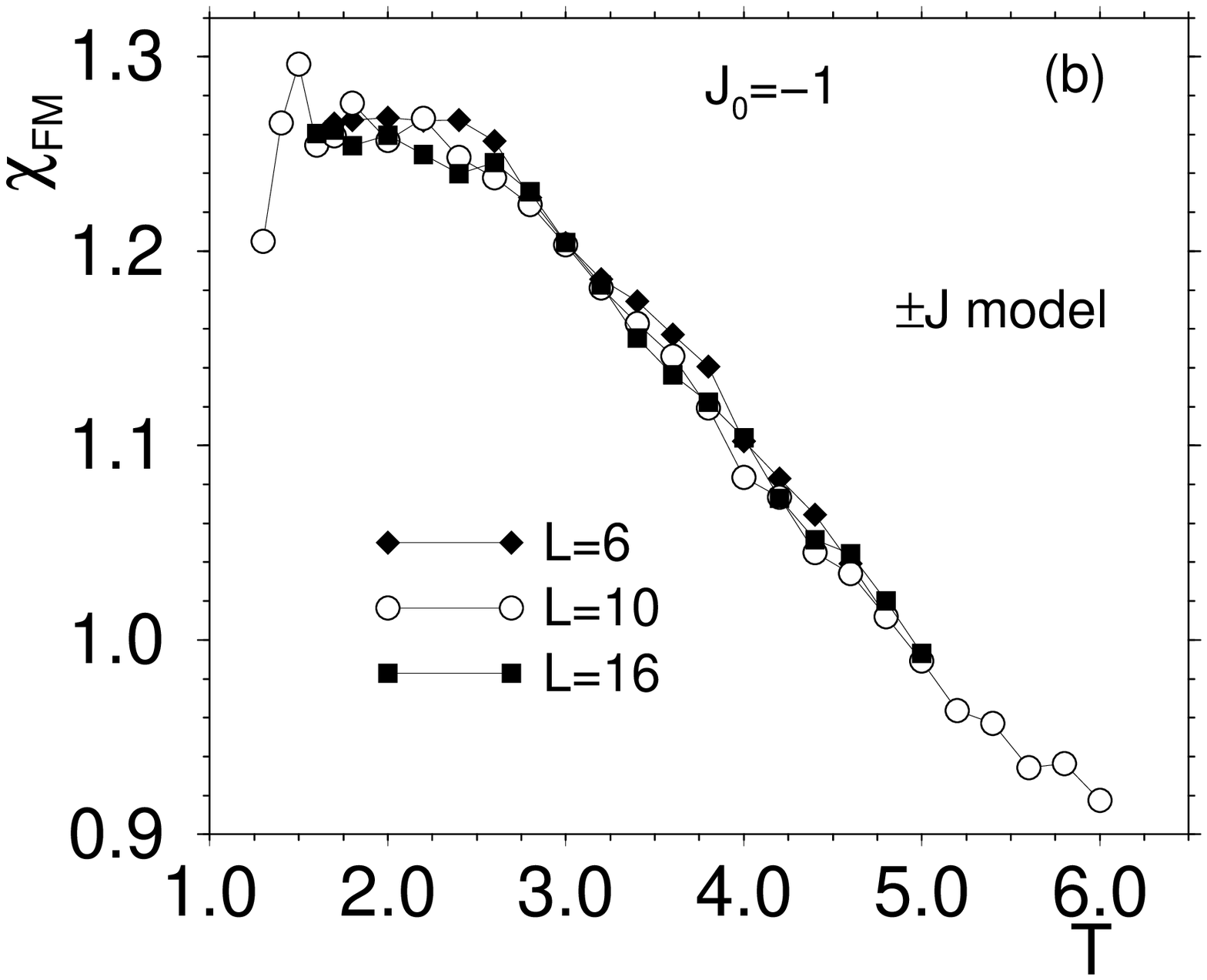,width=8.2cm,height=7.0cm}
}
\vspace*{-8mm}

\caption{\label{fig2} Temperature dependence of the ferromagnetic
susceptibility $\chi_{FM}=N[\langle m^2 \rangle ]_{\textrm{av}}$ in the
$p=10$ Potts glass for (a) $J_0=0$ and (b) $J_0=-1$. The three curves
correspond to different system sizes.}
\end{figure}

\begin{figure}
\centerline{
\psfig{figure=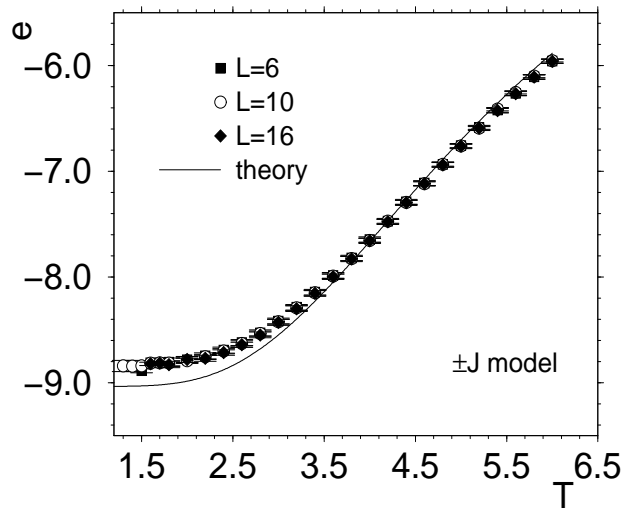,width=8.2cm,height=7.0cm}
}
\vspace*{-8mm}

\caption{\label{fig3} Temperature dependence of the energy per spin
$e(T)$ for three lattice sizes as indicated in the figure.  Note that
for $T \geq 1.5$ the average $[\cdots]_{\textrm{av}}$ was realized by
averaging over 100 independent realizations for $L=6$ and $L=10$, and by
averaging over 50 independent realizations for $L=16$. For $T=1.3$ and
$T=1.4$, only $L=10$ was studied, and only 10 realizations were used,
since 10 times longer runs than usual were necessary. The solid curve
is the theoretical prediction using an approach of independent clusters
(see main text for details).}
\end{figure}

\begin{figure}
\centerline{
\psfig{figure=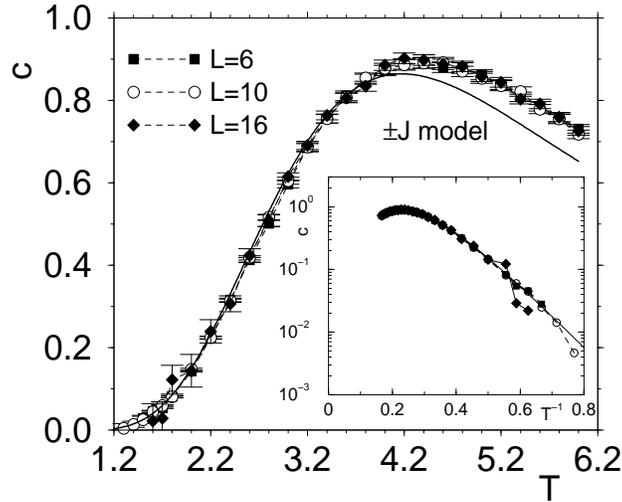,width=8.2cm,height=7.0cm}
}
\vspace*{-8mm}

\caption{\label{fig4} Specific heat per spin $c(T)$ of the bimodal Potts
glass plotted vs. temperature for three lattice sizes, as indicated in
the figure. The bold curve is the result of a theoretical calculation
using an approach of independent clusters (see main text for details).
The inset shows a lin-log plot of c(T) vs. $1/T$, in order to show
the $T-$dependence at low temperatures.}
\end{figure}

\begin{figure}
\centerline{
\psfig{figure=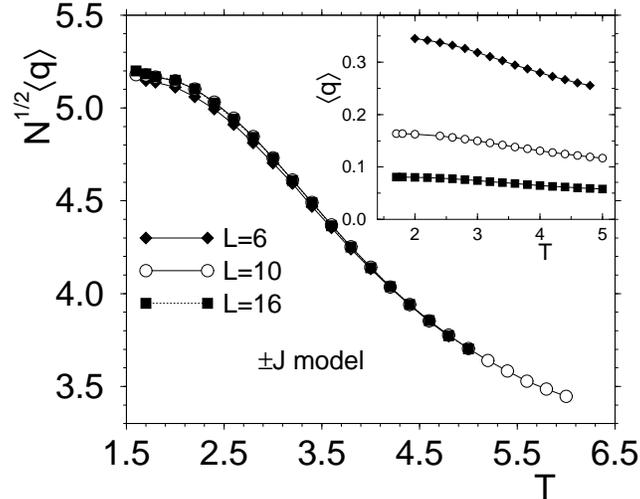,width=8.2cm,height=7.0cm}
}
\vspace*{-8mm}

\caption{\label{fig5} Scaled first moment $N^{1/2} \langle
q\rangle$ of the order parameter plotted as a function of
temperature. Different curves correspond to the different choices
of $L$, as indicated. The inset shows the $T-$dependence of $\langle
q\rangle$.}
\end{figure}

\begin{figure}
\centerline{
\psfig{figure=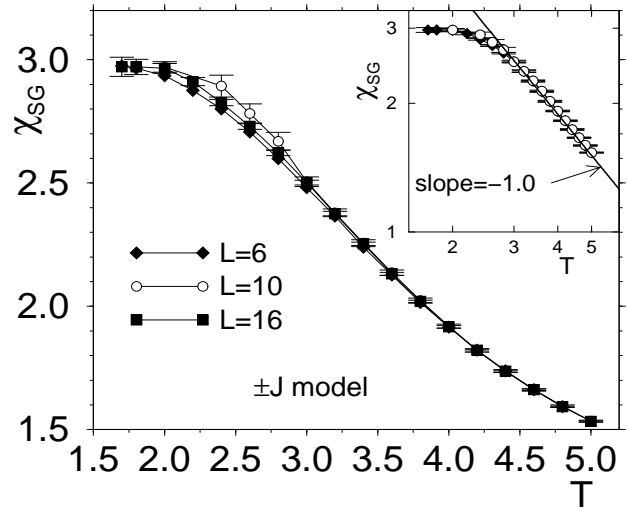,width=8.2cm,height=7.0cm}
}
\vspace*{-8mm}

\caption{\label{fig6} Spin glass susceptibility $\chi_{\rm SG}$
as defined in Eq.~(\ref{eq13}) as a function of temperature. The inset shows
the same data in a log-log plot, to demonstrate that at high $T$ we have 
$\chi_{\rm SG} \propto 1/T$.}
\end{figure}

\begin{figure}
\centerline{
\psfig{figure=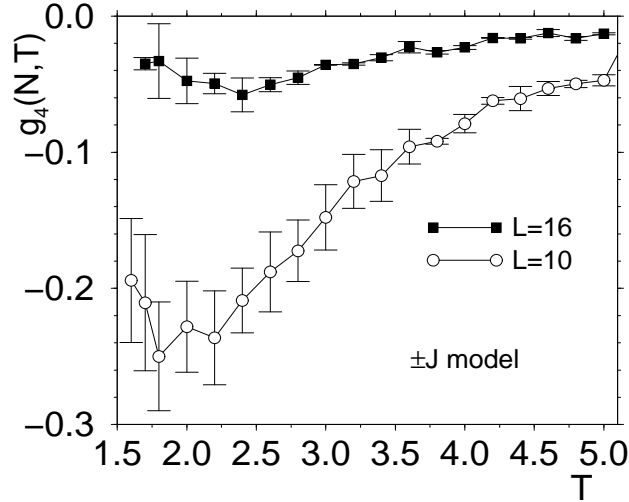,width=8.2cm,height=7.0cm}
}
\vspace*{-8mm}

\caption{\label{fig7} Plot of the reduced fourth-order cumulant
$g_4(N,T)$ versus temperature, for $L=10$ (open circles) and
$L=16$ (filled squares). The data for $L=6$ were rather noisy and
therefore were not included.}
\end{figure}

\begin{figure}
\centerline{
\psfig{figure=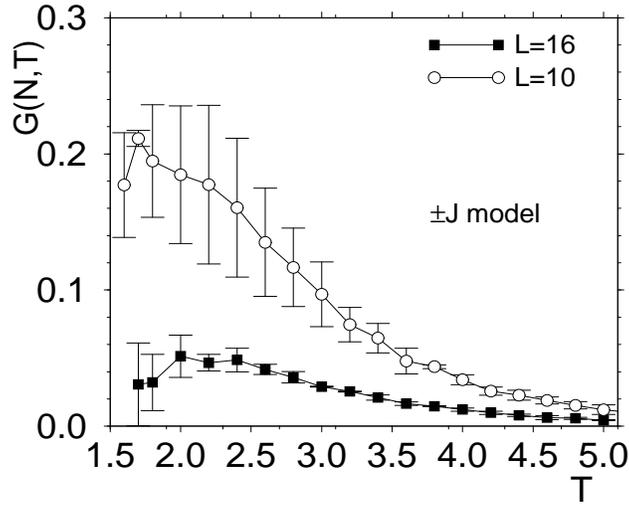,width=8.2cm,height=7.0cm}
}
\vspace*{-8mm}

\caption{\label{fig8} Plot of the Guerra parameter $G(N,T)$ versus
temperature for $L=10$ (open circles) and $L=16$ (filled squares).}
\end{figure}

\begin{figure}
\centerline{
\psfig{figure=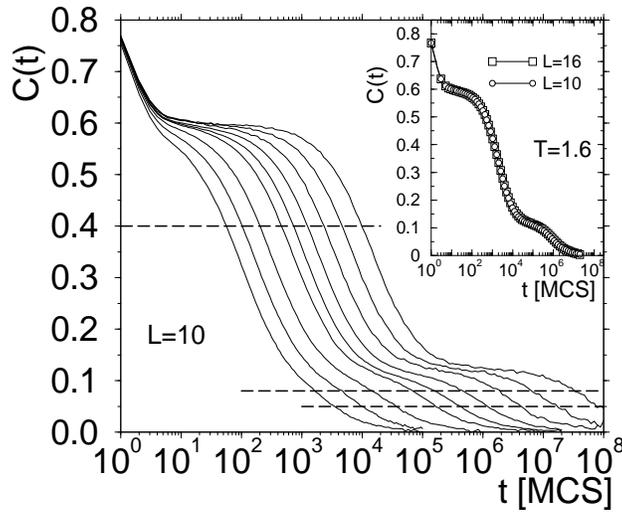,width=8.2cm,height=7.0cm}
}
\vspace*{-8mm}

\caption{\label{fig9} Time dependence of the spin auto-correlation
function of the $\pm J$ Potts glass for $L=10$ and various temperatures
(left to right: $T=2.4, 2.2, 2.0, 1.8,1.7,1.6, 1.5, 1.4$, and $1.3$). The
three horizontal dashed lines are used to define the relaxation times $\tau_i$
(see main text for details). In the inset we compare $C(t)$ for $L=10$
and $L=16$ at $T=1.6$.}
\end{figure}

\begin{figure}
\centerline{
\psfig{figure=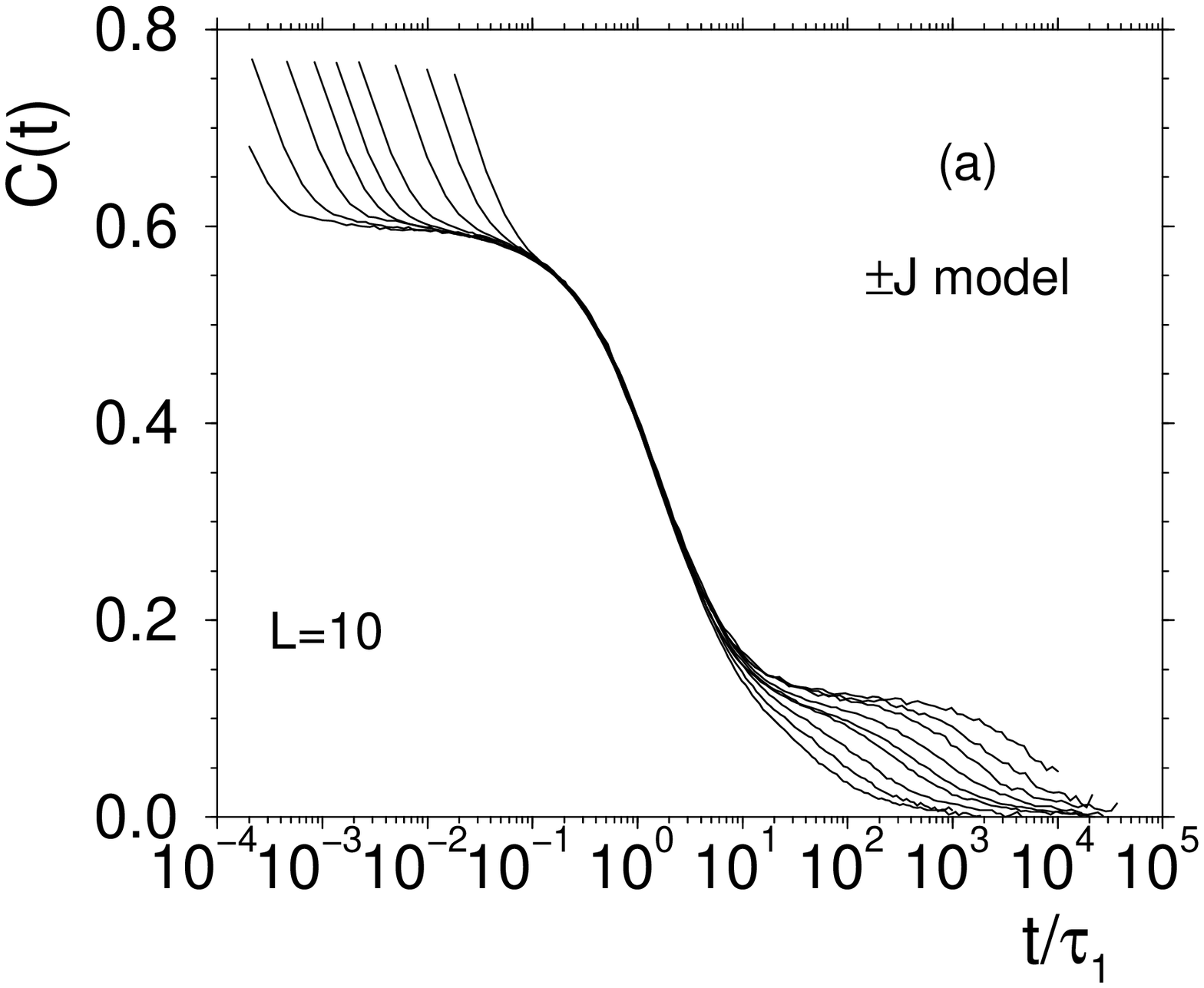,width=8.2cm,height=7.0cm}
\psfig{figure=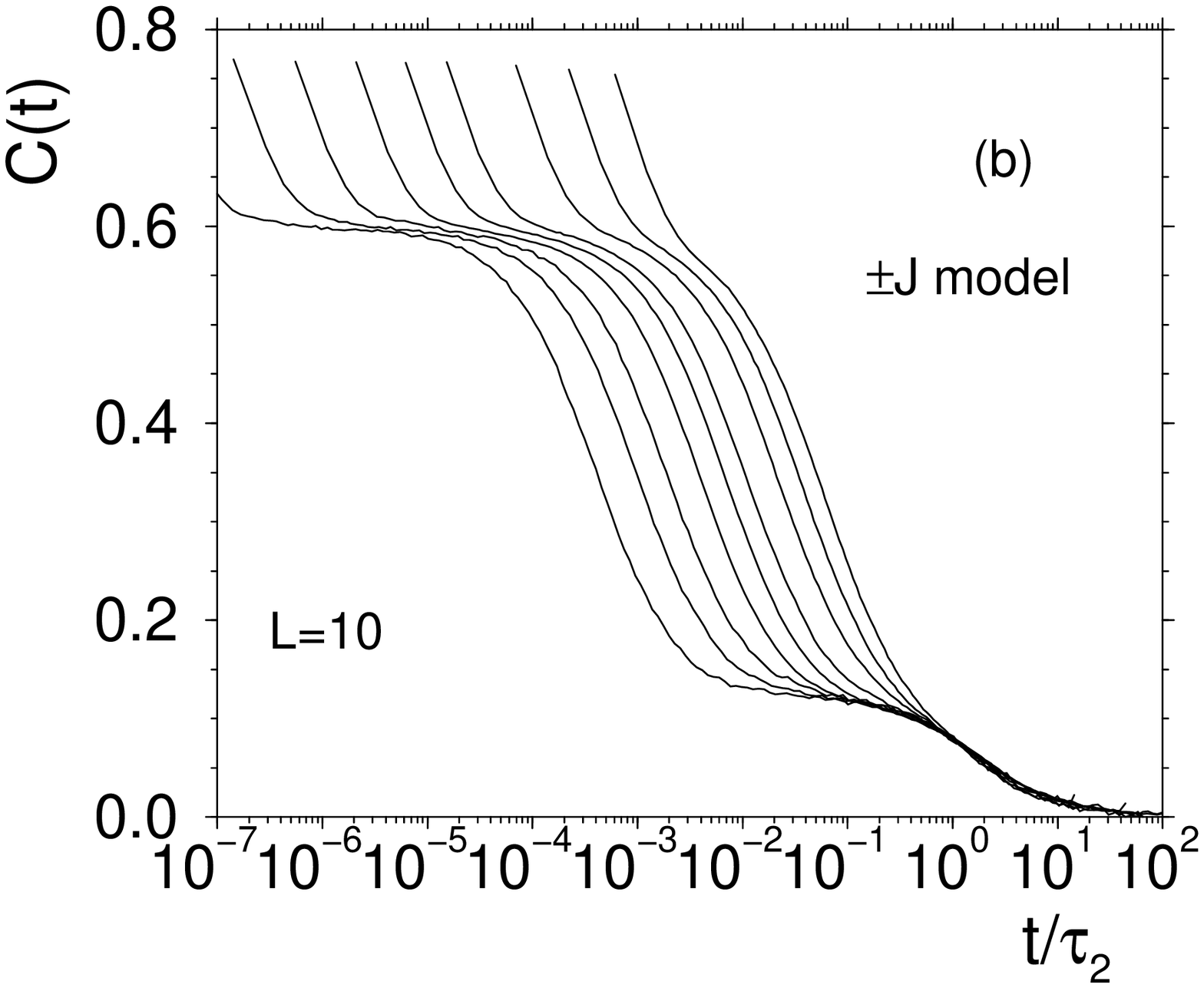,width=8.2cm,height=7.0cm}
}
\vspace*{-8mm}

\caption{\label{fig10} a) Spin auto-correlation function $C(t)$
for $L=10$ plotted vs. $t/\tau_1$. b) Spin auto-correlation
function $C(t)$ for $L=10$ plotted vs. $t/\tau_2$. The data for
$C(t)$ of this figure are identical to the ones shown in
Fig.~\ref{fig9}.}
\end{figure}

\begin{figure}
\centerline{
\psfig{figure=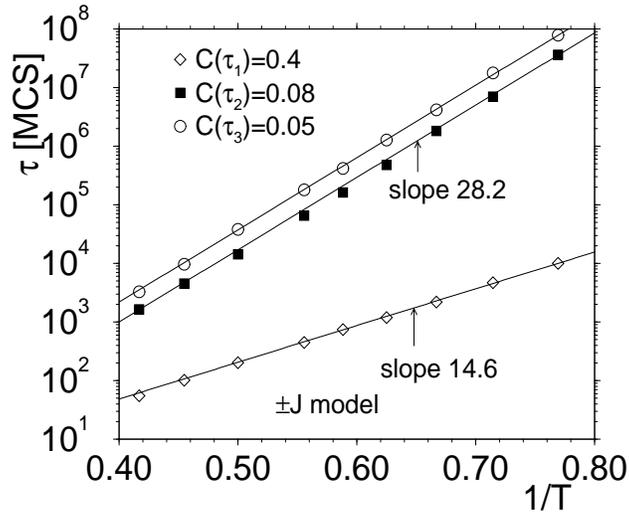,width=8.2cm,height=7.0cm}
}
\vspace*{-8mm}

\caption{\label{fig11} Arrhenius plot of the relaxation times $\tau_1$,
$\tau_2$, and $\tau_3$ defined in Eq.~(\ref{eq20}), for the $\pm J$
Potts glass. The straight lines are fits with an Arrhenius law and from
their slopes the activation energies $E_A^{(i)}$, see Eq.~(\ref{eq21}),
are estimated.}
\end{figure}

\begin{figure}
\centerline{
\psfig{figure=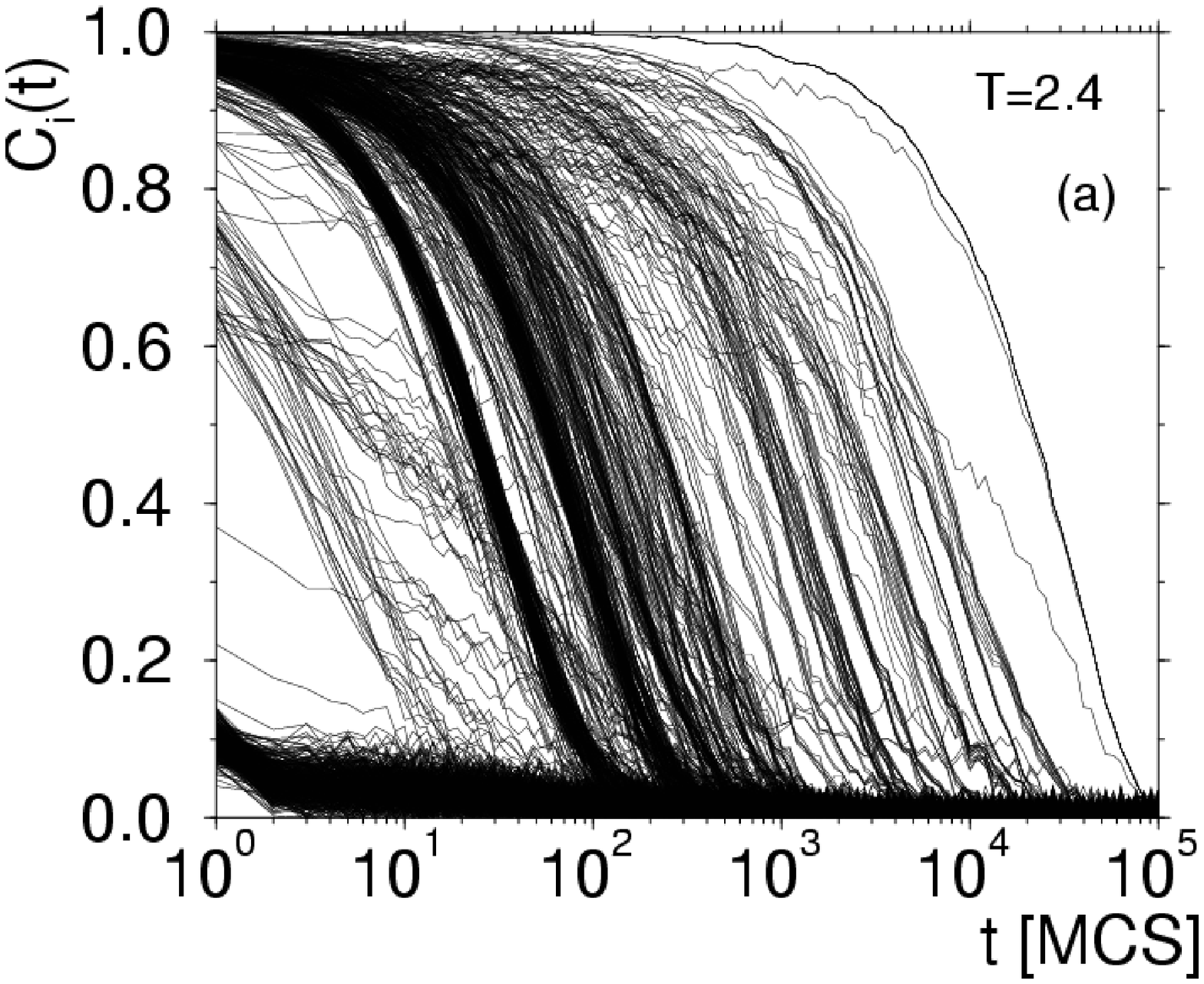,width=8.2cm,height=7.0cm}
\psfig{figure=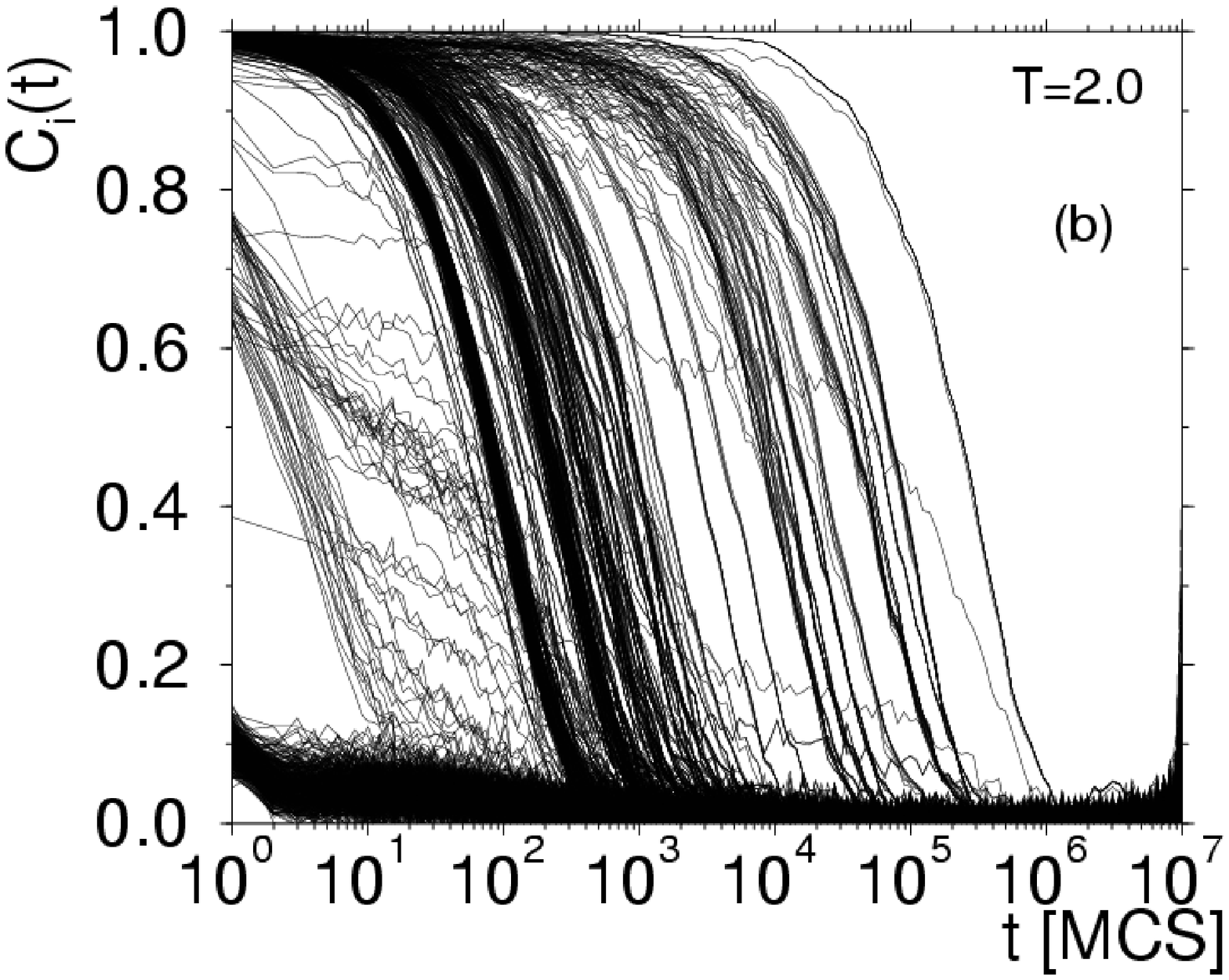,width=8.2cm,height=7.0cm}
}
\vspace*{-2mm}
\centerline{
\psfig{figure=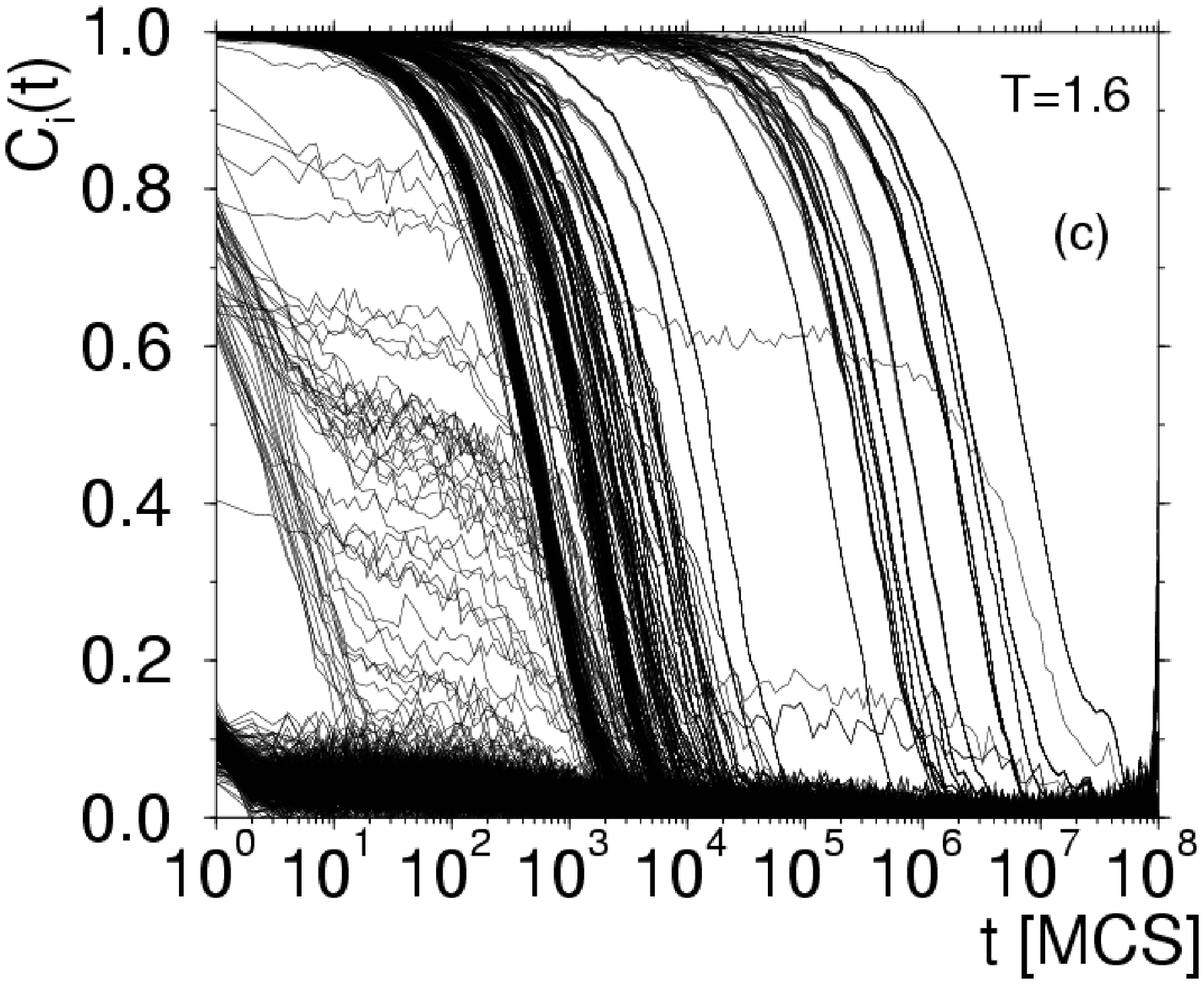,width=8.2cm,height=7.0cm}
}
\vspace*{-3mm}

\caption{\label{fig12} Time dependence of the relaxation function
$C_i(t)=(p-1)^{-1} \langle \vec{S}_i(t') \cdot \vec{S}_i(t' + t) \rangle$
of individual spins of the $\pm J$ Potts glass for $L=10$ at three different
temperatures: (a)~$T=2.4$, (b) $T=2.0$, and (c) $T=1.6$.}
\end{figure}

\begin{figure}
\centerline{
\psfig{figure=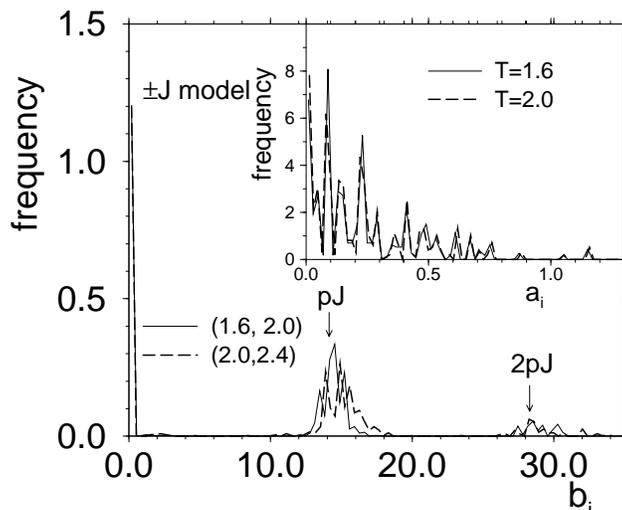,width=8.2cm,height=7.0cm}
}
\vspace*{-8mm}

\caption{\label{fig13} Normalized distribution of barrier heights
$b_i$ and pre-factors $a_i$ (inset) obtained from fitting the
Arrhenius relation, Eq.~(\ref{eq22}), to the relaxation times
$\tau_i'$ of individual spins, using either the two temperatures
$T=1.6, 2.0$ (full curves) or $T=2.0,2,4$ (dashed curves).}
\end{figure}

\begin{figure}
\centerline{
\psfig{figure=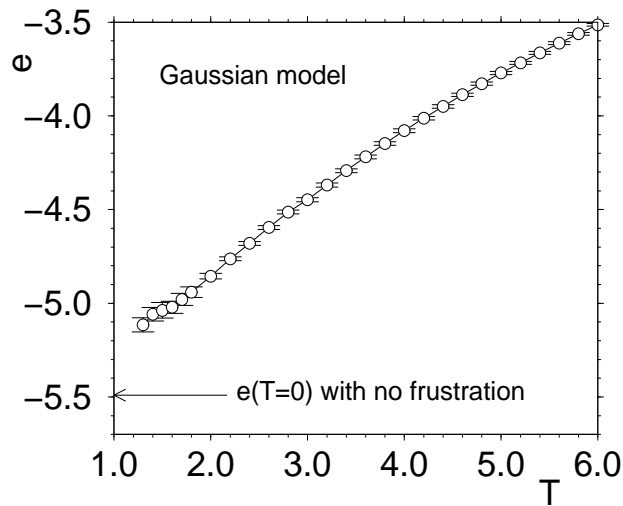,width=8.2cm,height=7.0cm}
}
\vspace*{-8mm}

\caption{\label{fig14} Energy per spin $e(T)$ of the $10$-state
Gaussian Potts glass plotted versus temperature for $L=10$. The
arrow indicates the estimate of the ground state energy under the
assumption that there are no frustration effects (see Eq.~(\ref{eq22b})).}
\end{figure}

\begin{figure}
\centerline{
\psfig{figure=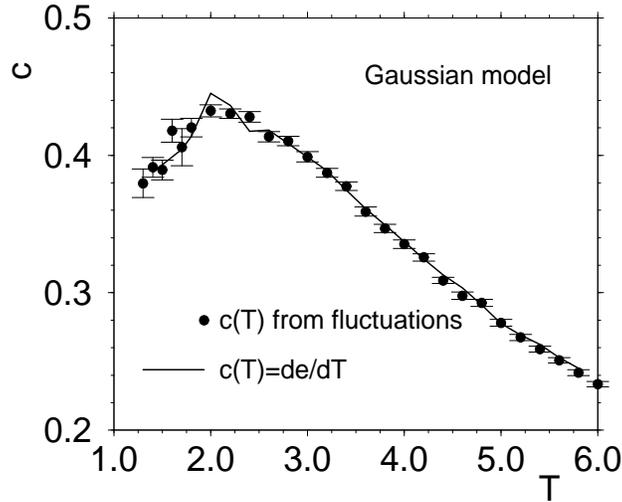,width=8.2cm,height=7.0cm}
}
\vspace*{-8mm}

\caption{\label{fig15} Specific heat $c(T)$ per spin of the
$10-$state Gaussian Potts glass plotted versus temperature for
$L=10$. Points are from energy fluctuations, while the broken line
is a numerical derivative of the curve $e(T)$ shown in
Fig.~\ref{fig14}.}
\end{figure}

\begin{figure}
\centerline{
\psfig{figure=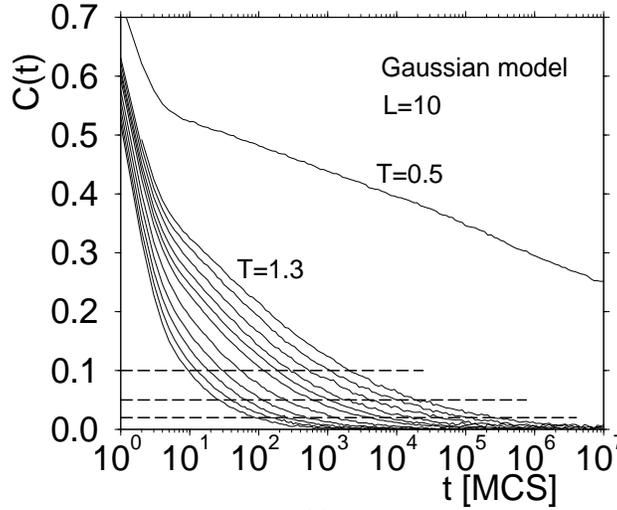,width=8.2cm,height=7.0cm}
}
\vspace*{-8mm}

\caption{\label{fig16} Spin auto-correlation function $C(t)$ of the
Gaussian Potts glass plotted versus time for $L=10$. The temperatures
are (from left to right): $T=2.8$, 2.6, 2.4, 2.2, 2.0, 1.8, 1.7, 1.6,
1.5, 1.4, 1.3 and 0.5. The horizontal dashed lines are used to define
the relaxation times $\tau_i$ given in Eq.~(\ref{eq24}). Note that the
data for $T=0.5$ is not in equilibrium (see main text for details).}
\end{figure}

\end{document}